\documentclass[aps,prl,reprint,floatfix,superscriptaddress]{revtex4-1}

\pdfoutput=1
\usepackage{graphicx}
\usepackage{color}
\usepackage{natbib}

\begin{document}


\title{3D integrated superconducting qubits}


\author{D. Rosenberg}
\email{drosenberg@ll.mit.edu}
\affiliation{MIT Lincoln Laboratory, 244 Wood Street, Lexington, MA 02420, USA}
\author{D. Kim}
\affiliation{MIT Lincoln Laboratory, 244 Wood Street, Lexington, MA 02420, USA}
\author{R. Das}
\affiliation{MIT Lincoln Laboratory, 244 Wood Street, Lexington, MA 02420, USA}
\author{D. Yost}
\affiliation{MIT Lincoln Laboratory, 244 Wood Street, Lexington, MA 02420, USA}
\author{S. Gustavsson}
\affiliation{Research Laboratory of Electronics, Massachusetts Institute of Technology, Cambridge, MA 02139, USA}
\author{D. Hover}
\affiliation{MIT Lincoln Laboratory, 244 Wood Street, Lexington, MA 02420, USA}
\author{P. Krantz}
\affiliation{Research Laboratory of Electronics, Massachusetts Institute of Technology, Cambridge, MA 02139, USA}
\author{A. Melville}
\affiliation{MIT Lincoln Laboratory, 244 Wood Street, Lexington, MA 02420, USA}
\author{L. Racz}
\affiliation{MIT Lincoln Laboratory, 244 Wood Street, Lexington, MA 02420, USA}
\author{G. O. Samach}
\affiliation{MIT Lincoln Laboratory, 244 Wood Street, Lexington, MA 02420, USA}
\author{S. J. Weber}
\affiliation{MIT Lincoln Laboratory, 244 Wood Street, Lexington, MA 02420, USA}
\author{F. Yan}
\affiliation{Research Laboratory of Electronics, Massachusetts Institute of Technology, Cambridge, MA 02139, USA}
\author{J. Yoder}
\affiliation{MIT Lincoln Laboratory, 244 Wood Street, Lexington, MA 02420, USA}
\author{A. J. Kerman}
\affiliation{MIT Lincoln Laboratory, 244 Wood Street, Lexington, MA 02420, USA}
\author{W. D. Oliver}
\affiliation{MIT Lincoln Laboratory, 244 Wood Street, Lexington, MA 02420, USA}
\affiliation{Research Laboratory of Electronics, Massachusetts Institute of Technology, Cambridge, MA 02139, USA}
\affiliation{Department of Physics, Massachusetts Institute of Technology, Cambridge, MA 02139, USA}



\date{\today}

\begin{abstract}
As the field of superconducting quantum computing advances from the few-qubit stage to larger-scale processors, qubit addressability and extensibility will necessitate the use of 3D integration and packaging. While 3D integration is well-developed for commercial electronics, relatively little work has been performed to determine its compatibility with high-coherence solid-state qubits. Of particular concern, qubit coherence times can be suppressed by the requisite processing steps and close proximity of another chip. In this work, we use a flip-chip process to bond a chip with superconducting flux qubits to another chip containing structures for qubit readout and control. We demonstrate that high qubit coherence ($T_1$, $T_{2,\rm{echo}} > 20\,\mu$s) is maintained in a flip-chip geometry in the presence of galvanic, capacitive, and inductive coupling between the chips.
\end{abstract}

\maketitle


Superconducting qubits are a prime candidate for constructing large-scale quantum processors due to their lithographic scalability, compatibility with microwave control, gate speed, and relatively long coherence times in planar geometries \cite{oliver_welander_2013,Devoret_2013}. Recent increases in coherence times \cite{Barends_2013, Rigetti_2012, Yan_2016} and the development of fast, high-fidelity single- \cite{Sheldon_2016,Barends_2014,Rol_2016} and two-qubit- gates \cite{Barends_2014, Sheldon_2016_2} have yielded control fidelities that exceed the most lenient thresholds required for fault tolerant quantum error correction via the surface code \cite{Fowler_2012}, a code of particular interest because it requires only nearest-neighbor interactions between qubits. With this motivation, recent experiments have prototyped basic error-detection codes, Bell-state memories, and multi-qubit entangled states using four \cite{C_rcoles_2015}, five \cite{Rist__2015}, nine \cite{Kelly_2015}, and ten qubits \cite{Song_2017}  in a planar geometry. While these experiments are important demonstrations of the underlying qubit technology, the devices were all controlled and read out using interconnects that laterally addressed the qubits from the perimeter of the same chip. Extending this approach to larger numbers of qubits is impractical due to the interconnect crowding that occurs when addressing qubits within a large two-dimensional array. Moving into the third dimension eases such geometrical constraints, enabling efficient interconnect routing to large 2D arrays, allowing for more compact qubit-qubit coupling geometries, and affording significantly increased connectivity – beyond nearest-neighbor interactions – that is advantageous for many  error correcting codes \cite{Bombin_2006,Fowler_2012,Kovalev_2013} and of importance to quantum annealing and quantum simulation.

One method for accessing the third dimension is to use monolithic fabrication techniques to create a planarized multi-layer structure. This method has been used in the D-Wave quantum annealing processors containing up to $2000$ qubits \cite{dwave}. However, with current fabrication techniques the price of monolithic fabrication is a severe penalty on qubit coherence, as evidenced by the low coherence time of the qubits in the D-Wave processor compared with state-of-art in single-layer aluminum devices \cite{Yan_2016, Barends_2013, Weber_2017}. A previous experiment used a flip-chip architecture with large sapphire spheres setting the spacing between two chips \cite{Li_2010}, but the assembly method used was not scalable and lacked galvanic connection between the chips. More recent efforts have focused on scalable vertical interconnects \cite{B_janin_2016, Versluis_2016,Liu_2017}, but these approaches have not yet demonstrated compatibility with high-coherence superconducting qubits.

\begin{figure}[h]
\includegraphics[width = \columnwidth]{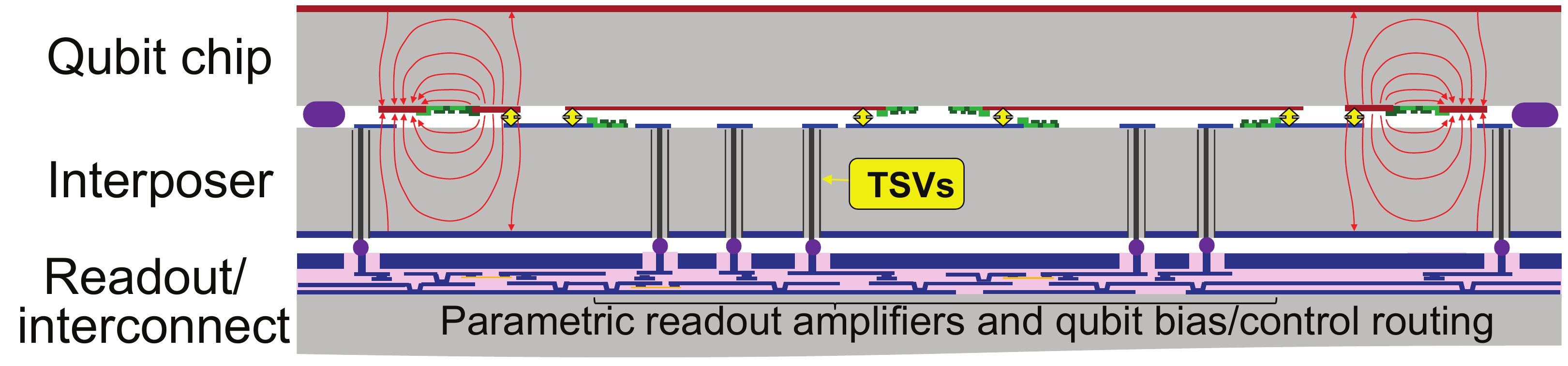}%
\caption{\label{fig:ThreeStack}  Envisioned scheme for control and readout of a large-scale, 3D integrated quantum processor. The qubit, interposer, and readout/interconnect chips are connected using indium bump bonds. The qubits are separated from the readout and control layer by an interposer chip with through-substrate vias that provide input/output (I/O) connectivity to/from the qubits. Because the chips are fabricated separately, each fabrication process can be optimized independently.}
\end{figure}

Here we describe a new approach that leverages heterogeneous 3D integration to create an architecture that enables use of the third dimension without sacrificing qubit performance, and we present proof-of-principle experimental data indicating the feasibility of this approach. Fig. \ref{fig:ThreeStack} shows a schematic of our envisioned structure. The design consists of three chips, attached using superconducting bump bonds, with each chip performing a different function. The top chip contains the superconducting qubits that are the basic logic elements of the quantum processor. The middle interposer chip has patterned surfaces on both sides of the chip, with metalized through-silicon vias (TSVs) providing connectivity between the two surfaces. The bottom chip uses a multilayer planarized process for efficient wire routing \cite{Tolpygo_2015} and active Josephson junctions for signal amplification \cite{Macklin_2015}. In this design, elements on the top surface of the interposer chip, close to the qubit, are galvanically, inductively, or capacitively coupled to the qubits for bias, control, and readout, and these elements connect to the signal readout and interconnect chip through the TSVs and the indium bumps. This design has two significant advantages. First, the fabrication processes for each chip can be performed separately and independently. This is particularly an advantage for fabrication of the qubits, which are notoriously sensitive to materials and processes that can cause decoherence \cite{oliver_welander_2013}. Second, the thick interposer chip provides a large mode volume for the qubit electromagnetic fields as well as isolation between the qubit and interconnect/readout chip, ensuring that the qubit performance is not degraded by the added system complexity.

\begin{figure*}[htb]
\includegraphics[width=\textwidth]{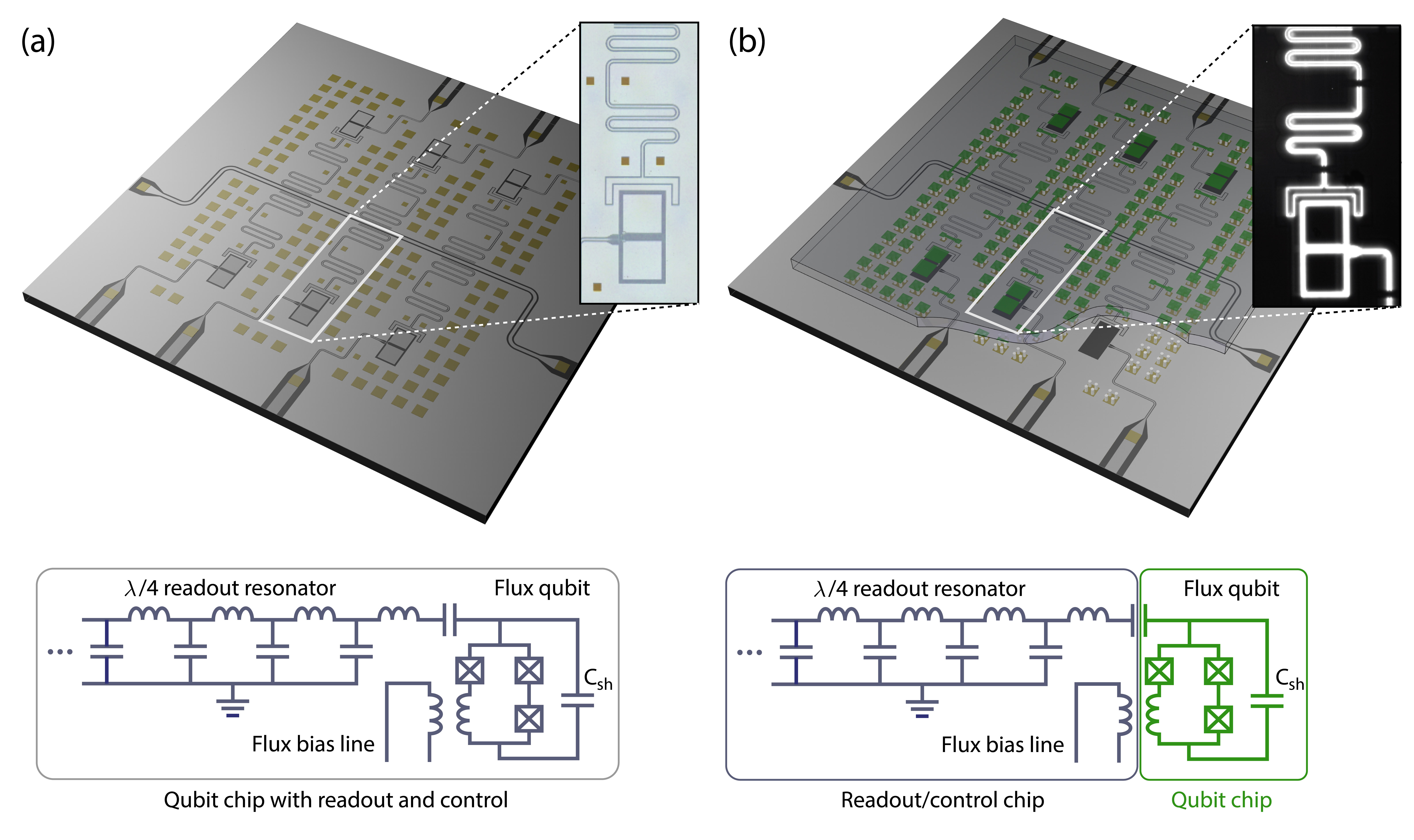}
\caption{\label{fig:ChipSchematic} Standard (a) and flipped qubit chip (b) configurations. a) Schematic of standard qubit chip with six capacitively-shunted flux qubits. Each qubit, which comprises a loop with three Josephson junctions shunted by a large capacitor, is capacitively coupled to a quarter-wave resonator for dispersive readout and control, and inductively coupled to a flux bias line. In this configuration, all readout and control elements are on the qubit chip. The array of small squares are the under bump metallization layer. An optical micrograph of one of the qubits and its corresponding readout resonator is shown to the right. b) Schematic of a flip-chip qubit chip. In this configuration, the qubits are on one chip, whereas the readout and control elements are on another chip that is bonded to the qubit chip. For visibility, the metal on the qubit chip is shown in green in the schematic and on the circuit diagram. An infrared through-chip image of one of the qubits and readout resonators is shown to the right.}
\end{figure*}

A first step towards assessing the practicality of the 3D integration illustrated in Fig. \ref{fig:ThreeStack} is to determine its impact on qubit performance. The presence of an additional surface proximate to the qubit may introduce new sources of noise, reducing qubit coherence time. In addition, 3D integration generally requires additional processing steps, such as depositing additional metal layers and bonding the chips, that may affect qubit performance. To quantify the effect of 3D integration on the qubit, we performed experiments using an intermediate architecture where a qubit chip is bonded to a single chip –- an interposer without TSVs -– using indium bumps. This allows us to determine the impact of 3D integration and to demonstrate basic desirable functionalities enabled by 3D integration, such as off-chip control and readout of the qubit.

For the experiments described here, we fabricated capacitively-shunted aluminum flux qubits on 2" silicon wafers using a process that has been described elsewhere \cite{Yan_2016}. As shown in Fig. \ref{fig:ChipSchematic}a, each chip contains six qubits, each of which is inductively coupled to a bias line for applying flux to shift the qubit energy levels and capacitively coupled to a quarter-wave resonator for control and dispersive readout. The qubits have relatively large loop areas, a design choice related to their application to quantum annealing, and  generally have $T_1,T_2 \approx 20\,\mu$s, somewhat lower than obtained for gate-based smaller-loop designs with $T_1,T_2 \approx 50\,\mu$s \cite{Yan_2016}.

 The resonators, which are spectrally spaced by approximately 50 MHz, are all inductively coupled to a transmission line for multiplexed readout and control. Our bump bonding approach included the addition of a patterned under bump metalization (UBM) layer, a metal stack comprising Ti/Pt/Au, to our standard qubit fabrication process, in order to make contact to the aluminum and to provide a diffusion barrier to avoid the formation of intermetallic compounds. 

\begin{figure}[htb]
\includegraphics[width = \columnwidth]{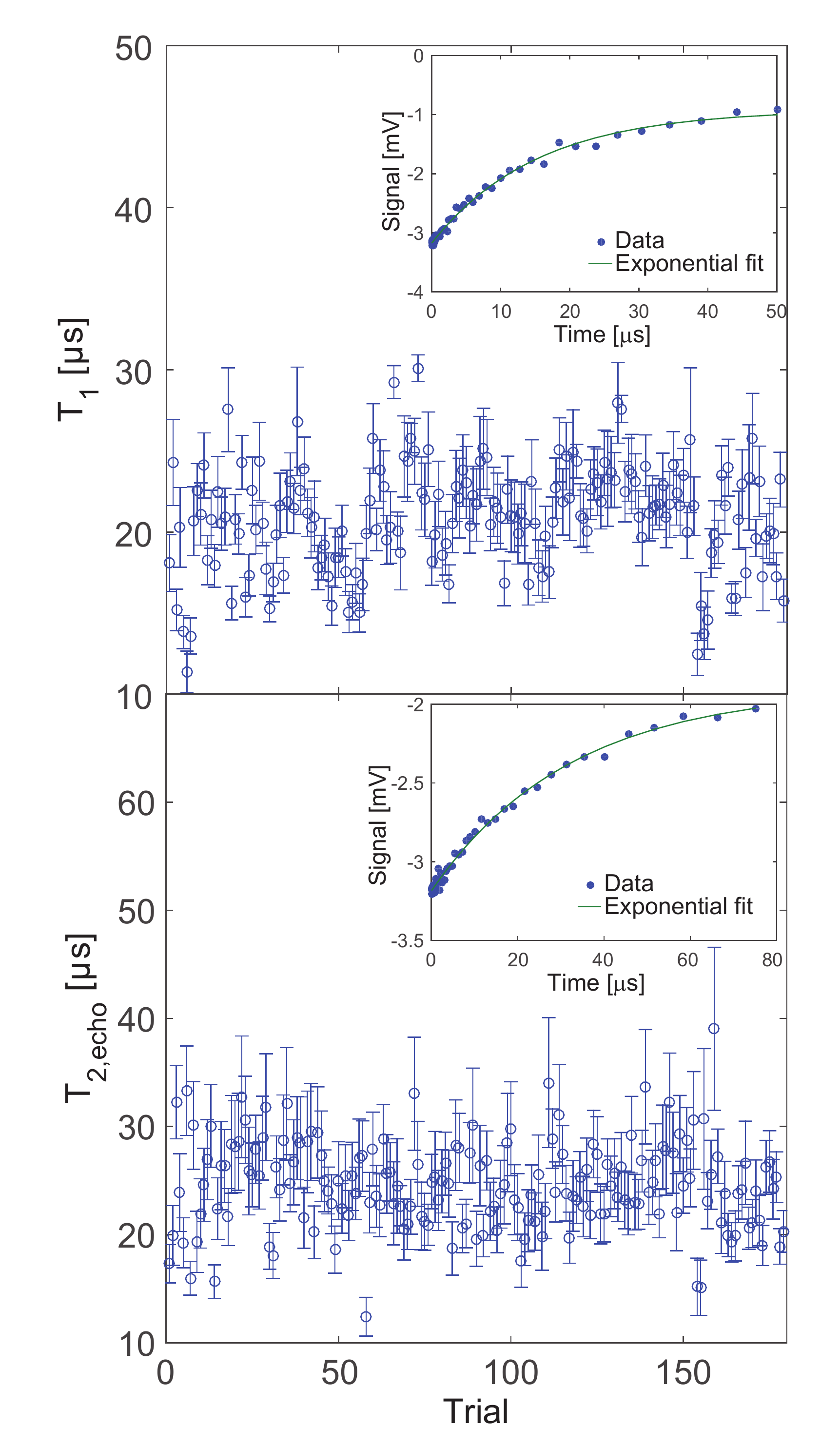}%
\caption{\label{fig:T1T2} T$_1$ and T$_{2, \rm{echo}}$ times for the flip-chip qubit configuration shown in Fig. \ref{fig:ChipSchematic}b, with a $10\,\mu$m standoff distance between the qubit and readout/control chips. Repeated interleaved scans were taken over the course of approximately 27 hours. The insets show single representative measurements of T$_1$ (T$_{2, \rm{echo}}$), and the main plots display the results of fits to an exponential decay curve with 95\% confidence error bars. The measured relaxation and coherence times are approximately equal to those for the experimental control shown in Fig. \ref{fig:ChipSchematic}a, where the qubit, readout, and control elements are all on one chip. }
\end{figure}

Qubits were designed to be tested either on stand-alone single chips, as shown in Fig. \ref{fig:ChipSchematic}a, or in a flip-chip configuration with separate qubit and control/readout chips, as shown in Fig. \ref{fig:ChipSchematic}b. For the flip-chip configuration, we bonded the qubit chip to a silicon interposer chip that contained structures (e.g., capacitors, inductors, transmission lines, etc.) patterned from a 250\,nm layer of evaporated aluminum. As with the qubit wafers, a patterned under bump metallization layer was used for making electrical contact to the aluminum. Thick (8-30$\,\mu$m tall) pillars of indium with diameters of 15 and 30 $\mu$m were evaporated on top of the underbump metal (squares in Fig. \ref{fig:ChipSchematic}), and a commercial thermo-compression bonder was used in force- or distance-feedback mode to bond the chips together. Three-dimensional images of the two bonded chips using a white-light interferometer and a confocal microscope indicated that the tilt angle between the two chips was less than $0.4\,$mRad, and infrared images showed an in-plane alignment error of less than $1\,\mu$m.

The design shown in Fig. \ref{fig:ChipSchematic}b is notable, both because the qubit chip is bonded to another chip and because all the structures used to control and read out the qubits are on the other chip. With the exception of the underbump metallization structures required for bump bonding and jumpers used to connect disparate sections of ground plane and improve the qubit microwave environment, the only structures remaining on the qubit chip are the qubits themselves. As shown in the color-coded schematic at the bottom of Fig. \ref{fig:ChipSchematic}, the flux bias line and readout resonator elements have been relocated to the control/readout chip and are inductively and capacitively coupled to the qubit across the gap separating the two chips. 

To the extent possible, we ensured that the flip-chip qubits were nominally the same as the planar qubits, which served as our experimental controls. This required design modifications in order to account for the vertical spacing ($2-10\,\mu$m) between the chips and the change in capacitance due to an increased effective dielectric constant resulting from the presence of the extra silicon chip. We took these effects into account and designed the chips so that the shunt capacitance, the capacitive coupling between the qubit and the resonator, and the mutual inductance between the flux bias line and the qubit loop were the same for the qubit designs shown in Figs. \ref{fig:ChipSchematic}a and \ref{fig:ChipSchematic}b. The chip in Fig. \ref{fig:ChipSchematic}b includes designs for four chip-to-chip spacings of $2\,\mu$m, $5\,\mu$m, $10\,\mu$m, and $20\,\mu$m. Simulations indicate that our design is fairly robust to deviations in the qubit-interposer spacing; for the $10\,\mu$m target design, a deviation of $1\,\mu$m results in a change in shunt and coupling capacitance of around 2\% and 5\%, respectively. In practice, we control the chip-to-chip spacing across the 5 mm x 5 mm chips to better than $1\,\mu$m.
 
To determine the impact of bump bonding on qubit coherence, we first tested the capacitively-shunted flux qubits in the standard single-chip configuration (Fig. \ref{fig:ChipSchematic}a). Based on noise spectroscopy measurements across a range of qubit designs fabricated using the same process \cite{Yan_2016}, it is expected that both charge and flux noise play a role in limiting the T$_1$ of these devices. As a result, our measurements are sensitive to increases in both charge and flux noise. The T$_1$ and T$_{2,\rm{echo}}$ times for these qubits were measured to be $10-20\,\mu$s, in reasonable agreement with measured flux and charge spectral noise densities \cite{Yan_2016}. We attribute the variance in the measurement to quasiparticle number variations \cite{Gustavsson1573}. 

Fig. \ref{fig:T1T2} shows measurements of T$_1$ and T$_{2,\rm{echo}}$ for the flip-chip qubit illustrated in Fig. \ref{fig:ChipSchematic}b, where the qubit is biased, controlled, and read out using structures on a separate chip at a standoff distance of $10\,\mu$m. We find that the average relaxation and echo times of $20.9\,\mu$s and $24.6\,\mu$s, respectively, are within the same range as those measured on qubits in the standard configuration (Fig. \ref{fig:ChipSchematic}a), indicating that 3D integration did not adversely affect the qubit. The slight increase in T$_1$ and T$_{2,\rm{echo}}$ compared to our control could be due to reduced participation of the surfaces in the qubits electric field \cite{Wang_2015}, but the increase is within the range of variations we generally observe in similar qubits.

\begin{figure*}[htb]
\includegraphics[width = \textwidth]{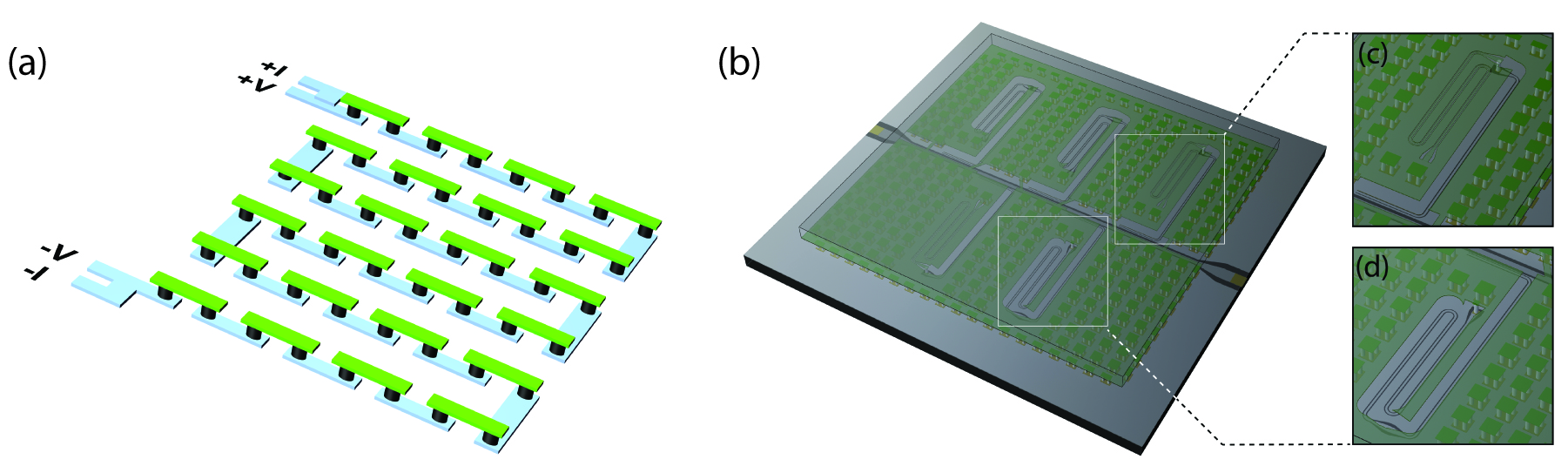}
\caption{\label{fig:ResStruc} Structures for measuring the DC and microwave properties of the bump interconnects. a) Representative schematic of a bump chain for measuring DC resistance. Strips of aluminum on two separate chips (shown in blue and green) are connected by indium bumps to form a continuous chain. The actual bump chains measured have 2,704 bump interconnects. b) Bump-interrupted quarter-wave resonators for measuring microwave loss. Coplanar waveguide resonators are split across two chips, with an indium bump providing connectivity between the two sections. The chips each have five resonators, one experimental control with no bump interconnects and four resonators with bump interconnects, two near the middle of the resonator, and two near the voltage node of the resonator. c) Resonator with bump interconnect near the middle. d) Resonator with bump interconnect at the voltage node.}
\end{figure*}

Although the demonstration of off-chip readout and control was enabled by capacitive and inductive coupling alone, direct galvanic connection between bump-bonded chips is required for the full architecture shown in Fig. \ref{fig:ThreeStack}. Using structures such as those shown in Fig. \ref{fig:ResStruc}, we have measured the inter-chip resistance at low frequencies using chains of bumps and at microwave frequencies using resonators with bump interconnects. For the low-frequency measurements, we performed low-temperature four-wire measurements of the bump chains using both a commercial multimeter and a lock-in amplifier at frequencies ranging from $2\,$Hz to $200\,$Hz. We observe changes in resistance at $3\,$K and at $1\,$K, which we attribute to the indium and the aluminum going through their respective superconducting transitions. Using a chain of 2,704 indium bumps, we measured a DC resistance of $240\,\rm{n}\Omega$ per bump at temperatures well below $1\,$K, consistent with estimates of the normal state resistance of the under bump metallization layer. To reduce this resistance, we note that the UBM may be replaced by superconducting materials such as TiN.

\begin{figure}[htb]
\includegraphics[width = \columnwidth]{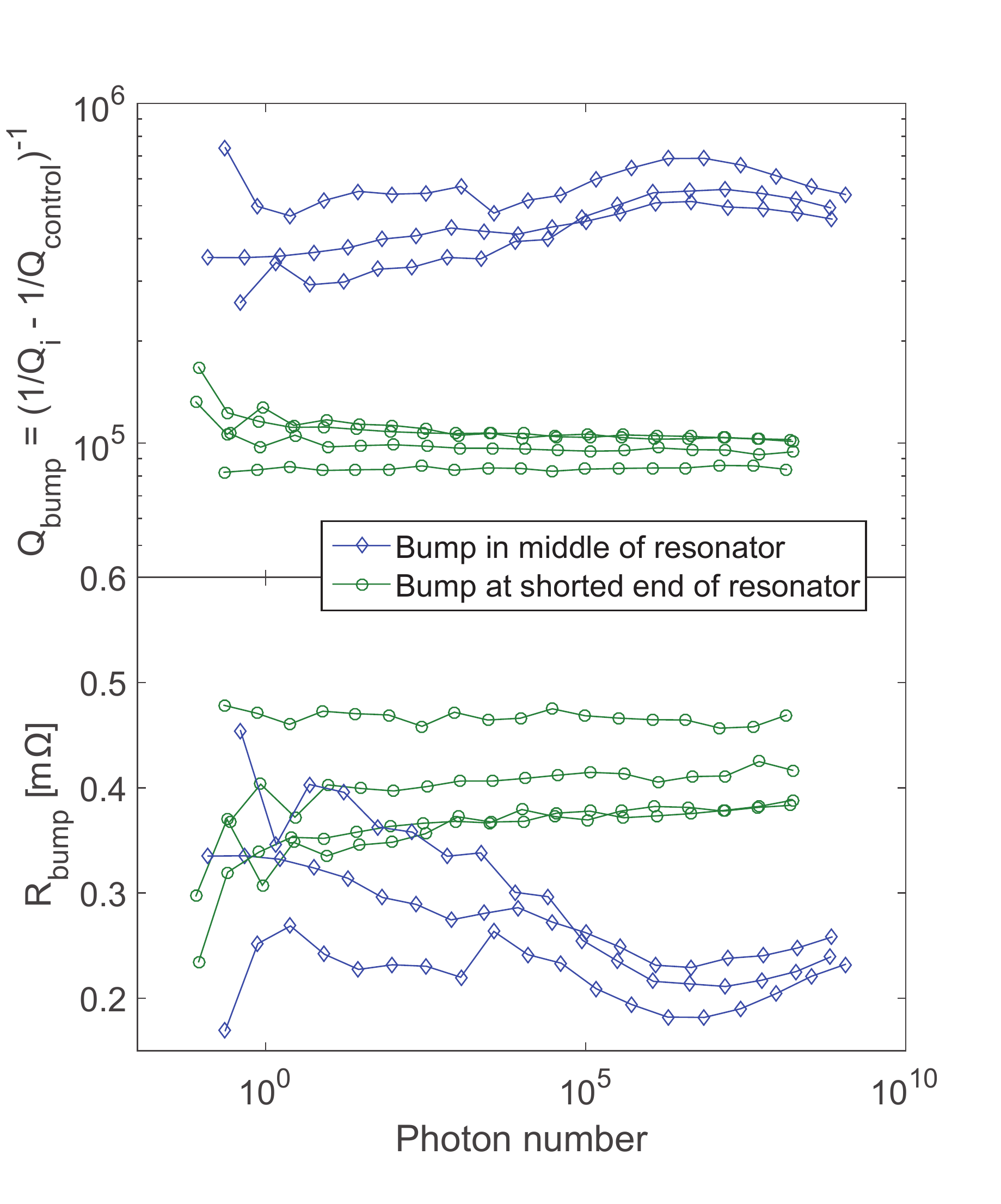}%
\caption{\label{fig:BumpRes} Measurements of the bump-interrupted resonators shown in Fig. \ref{fig:ResStruc}b. The top panel shows the contribution of the bump interconnect to the quality factor of the resonator, and the bottom panel shows the effective bump resistance, inferred by treating the bump as a series resistance.}
\end{figure}

In addition to DC signals, bumps will likely be used to pass microwave signals, e.g., for qubit control and readout. Therefore, it is important to quantify microwave loss due to the indium bumps, for example, from electrically-active two-level systems (TLS) on the bump surfaces that participate in the qubit electromagnetic mode volume. To measure the bump resistance at microwave frequencies, we designed quarter-wave transmission line resonators with bump interconnects, where the resistance of the bump is manifest as a reduction in the quality factor of the resonator. The resonators, with resonant frequencies ranging from $4.5\,$GHz to $5.5\,$GHz, were distributed between two chips, and a single bump with a $15\,\mu$m diameter (before compression) provided an electrical connection between the sections of the resonator. As shown in Fig. \ref{fig:ResStruc}b, we designed resonators where the bump interconnects were either at the shorted end of the resonator (the voltage node), or near the middle of the resonator, where both voltage and current oscillate at the resonant frequency. Depending on the dominant loss mechanism, we expect different results for the position dependence of the loss. If the loss mechanism is primarily through interaction with a bath of TLS, we expect the resonators with  interconnects near the middle of the resonator to exhibit a greater reduction in their quality factor, Q. If, on the other hand, the primary loss mechanism is resistive loss, we expect the resonators with interconnects at the shorted end of the resonator to have a greater Q reduction, since the interconnect is at the position with the greatest current flow. 

The top panel of Fig. \ref{fig:BumpRes} shows the reduction of quality factor for seven resonators across two chips. Four of the resonators had interconnects at the voltage node, and three had interconnects near the middle of the resonator. For each chip, the intrinsic quality factor due to other effects (e.g. material losses, vortices, radiation), $Q_{\rm{control}}$, was measured using a resonator with the same geometry but no bump interconnects. We then subtracted $Q_{\rm{control}}$ in parallel to obtain the Q reduction due to the bump $Q_{\rm{bump}} = (1/Q_i - 1/Q_{\rm{control}})^{-1}$. The resonators with the bumps at the voltage nodes clearly show a more pronounced reduction in Q, indicating that the bumps exhibit a series resistance at microwave frequencies. We extracted this resistance from our data by comparing the reduction in quality factor to simulations of an ideal coplanar waveguide resonator with varying resistance at the appropriate locations along the resonator. The extracted resistances are in the range of 0.1 to 0.5 m$\Omega$ (bottom panel of Fig. \ref{fig:BumpRes}). The slight power dependence seen in the resonators with bumps in the middle of the resonator is plausibly consistent with the behavior expected from two-level systems, which should saturate at high powers. If two-level systems were contributing to the Q-reduction, however, we would expect that at high photon numbers, where the TLS are saturated, the effective bump resistance would be equal to that obtained from the data with bumps at the voltage node. The inconsistency may be related to small systematic differences between the resonators,  but remains undetermined.

There are several factors which could contribute to the difference of three orders of magnitude between the DC and microwave resistance. First, the indium could intrinsically be lossier at microwave frequencies compared to DC frequencies. Second, the thin normal underbump layer could result in different current flows at DC and microwave frequencies. Finally, small differences in the design of the DC and microwave structures, such as the spacing of adjacent bumps, could contribute to the resistance difference. Overall, the data indicate that it should be possible to incorporate the bumps in transmission lines with microwave power levels suitable for qubit manipulation and measurement, but, in their present form, not in high-Q resonators or for transferring quantum information. A next step that may alleviate this issue is the use of superconducting underbump metals such as TiN.

Our demonstration of capacitive, inductive, and low-resistance galvanic coupling between two chips is a promising first step toward building larger-scale devices for quantum information processing. We have shown that it is possible to control and read out a qubit using off-chip elements while maintaining high qubit coherence. Although we have performed these initial demonstrations with flux qubits, designed for use in quantum annealers, these results are generally applicable to chip-based superconducting and semiconducting qubit modalities used for all forms of quantum information processing, including computation, annealing, and emulation. Additionally, for transmons and other qubits limited by surface dielectric losses, the enhanced capacitance provided by the flip-chip architecture enables the construction of smaller qubits with lower electric field surface participation. We believe these proof-of-principle experiments are the first step towards an architecture that will enable large-scale quantum processing with high-coherence qubits.


\begin{acknowledgments}
We gratefully acknowledge Jeffrey Birenbaum, Greg Calusine and Wayne Woods for useful discussions and M. Augeri. P. Baldo, G. Fitch, J. Lidell, K. Magoon, X. Miloshi, P. Murphy, B. Osadchy, A. Sevi, R. Slattery, C. Thoummaraj, D. Volfson and T. Weir for valuable technical assistance. This research was funded by the Office of the Director of National Intelligence (ODNI), Intelligence Advanced Research Projects Activity (IARPA) and by the Assistant Secretary of Defense for Research \& Engineering under Air Force Contract No. FA8721-05-C-0002. The views and conclusions contained herein are those of the authors and should not be interpreted as necessarily representing the official policies or endorsements, either expressed or implied, of ODNI, IARPA, or the US Government.
\end{acknowledgments}

\bibliography{FlipChipManuscript_arXiv_v2}

\end{document}